\documentclass[twocolumn]{aastex631}

\shorttitle{RY Scuti Chemical Abundances}
\shortauthors{Taft et al.}

\graphicspath{{./}{figures/}}

\usepackage{makecell}
\usepackage{natbib}
\setcitestyle{authoryear, open={(}, close={)}}

\usepackage{graphicx}
\usepackage{setspace}
\usepackage{threeparttable} 

\definecolor{darkgreen}{rgb}{0, 0.6, 0.15}
\usepackage{color,soul}

\newcommand{\Msun}{\mbox{$M_{\odot}$}}

\def\sun{\hbox{$\odot$}}
\def\earth{\hbox{$\oplus$}}

\def\farcs{\hbox{$.\!\!^{\prime\prime}$}}

\def\lesssim{\mathrel{\hbox{\rlap{\hbox{%
  \lower4pt\hbox{$\sim$}}}\hbox{$<$}}}}
\def\gtrsim{\mathrel{\hbox{\rlap{\hbox{%
  \lower4pt\hbox{$\sim$}}}\hbox{$>$}}}}

\def\farcs{\hbox{$.\!\!^{\prime\prime}$}}

\begin{document}

\title{Simulating Chemical Abundances in the Circumstellar Nebula of the Late Stage Binary RY Scuti}

\author[0000-0001-6482-4074]{Sarah H. Taft}
\affiliation{Minnesota Institute for Astrophysics, University of Minnesota, 116 Church Street, Minneapolis, MN 55455, USA}

\author{Robert D. Gehrz}
\affiliation{Minnesota Institute for Astrophysics, University of Minnesota, 116 Church Street, Minneapolis, MN 55455, USA}

\author{Charles E. Woodward}
\affiliation{Minnesota Institute for Astrophysics, University of Minnesota, 116 Church Street, Minneapolis, MN 55455, USA}

\author{Nathan Smith}
\affiliation{Steward Observatory, University of Arizona, 933 N. Cherry Avenue, Tucson, AZ 85721, USA}

\author{Isabelle Perron}
\affiliation{Minnesota Institute for Astrophysics, University of Minnesota, 116 Church Street, Minneapolis, MN 55455, USA}

\author{Annalisa Citro}
\affiliation{Minnesota Institute for Astrophysics, University of Minnesota, 116 Church Street, Minneapolis, MN 55455, USA}


\begin{abstract}

RY Scuti, thought to be a Wolf-Rayet (WR) progenitor, is a massive, 
post-main-sequence, binary star system undergoing Roche lobe 
overflow (RLOF). SOFIA (+FORCAST) spectroscopy of the inner, ionized 
region of RY Scuti's double ringed toroidal nebula  affirms the previous 
detection of the well-studied 12.81 $\mu$m \ion{Ne}{2} forbidden transition 
and reveals four distinct emission lines, including three previously 
undetected transitions, \ion{S}{3}, \ion{Fe}{3}, and \ion{S}{3}. Cloudy photoionization modeling of the four neon, sulfur, and iron lines was performed 
to derive fractional abundances (log($\frac{X}{H}$)) 
of \ neon \ at -3.19 $\pm$ 0.0251, sulfur at -3.76 $\pm$ 0.0487, and iron at -2.23 $\pm$ 0.0286. All three species are overabundant with respect to fiducial solar chemical abundances, especially iron. Our analysis suggests that the outer envelope of the primary star in the RY Scuti system is being stripped away via RLOF, leaving helium-rich and hydrogen-poor material visible to observation. This material also exhibits elevated neon, sulfur, and iron fractional abundances, consistent with RY Scuti evolving toward a WR object.

\end{abstract}



\section{Introduction} \label{sec:intro}


RY Scuti (HD 169515) is a massive, post-main-sequence, eclipsing binary 
system. The first detection of the forbidden \ion{Fe}{3} transition \citep{edlen39} was in RY Scuti. Located at a distance of 1.8 $\pm$ 0.1 kpc \citep{king79, smith01, smith02}, RY Scuti's eclipsing light curve indicates that it is undergoing late-stage RLOF (\citealt{antokhina88}; \citealt{djura01}; \citealt{melikian10}). The primary/donor star, the star whose mass is being stripped during RLOF, is a supergiant with spectral classification O9.7 Ibpe var \citep{walborn82}. The secondary/receiving/companion star, the star that is accreting mass from the donor star, is likely an O5 star \citep{smith11}. Mass calculations for each stellar 
component of the system vary, with earliest studies reporting a combined mass over 
100M$_{\odot}$ \citep{popper43}, while recent studies typically cite a primary (donor) mass of $\simeq$ 7-10~\Msun{} and secondary (companion) mass of 
$\simeq$ 25-30~\Msun{} \citep{skulskii92, sahade02, grundstorm07}. Prior to RY Scuti's current 
RLOF epoch, the primary star was likely the more massive star in the 
binary, having lost a significant portion of its mass to gravitational transfer \citep{smith11}.

RY Scuti is also known to host a compact, double-ringed, expanding, toroidal circumstellar nebula. The nebula is spatially resolved in images across multiple wavelengths, including radio with the Very Large Array (VLA) \citep{gehrz95}, infrared with Keck I \citep{gehrz01} and Keck II \citep{smith11}, and optical H$\alpha$ and [\ion{Si}{3}] with the Hubble Space Telescope (HST) \citep{smith99, smith01, smith02}. The inner ring of the toroidal nebula is $\simeq$ 2800~au in diameter. H$\alpha$ and [\ion{Si}{3}] HST images map out strong recombination lines from hot, ionized gas, and radio images map out free-free radiation from the same gas \citep{gehrz01}. The outer ring of the nebula is $\simeq$ 3800 au in diameter, and thermal emission reveals that it is primarily comprised of warm dust, potentially with a polycyclic aromatic hydrocarbon (PAH) component \citep{gehrz95, smith11}. The outer ring was ejected in 1754 $\pm$ 36 yr, and the inner, ionized ring in 1881 $\pm$ 4 yr \citep{smith01}. This subsequent ejection of what we now see as the inner, ionized ring allowed dust to form in the outer shell as the inner shell shielded the outer shell from hot, ionizing photons. The distance to the inner ring's boundary is likely consistent with the system's Str\"omgren radius \citep{gehrz95, smith99}. It is in this ionized region of RY Scuti's nebula where observed IR forbidden emission lines, such as the well-studied 12.81 $\mu$m \ion{Ne}{2} line \citep{gehrz95}, originate. 

\begin{figure*}[!hbt]
    \centering
    \includegraphics[width=6in]{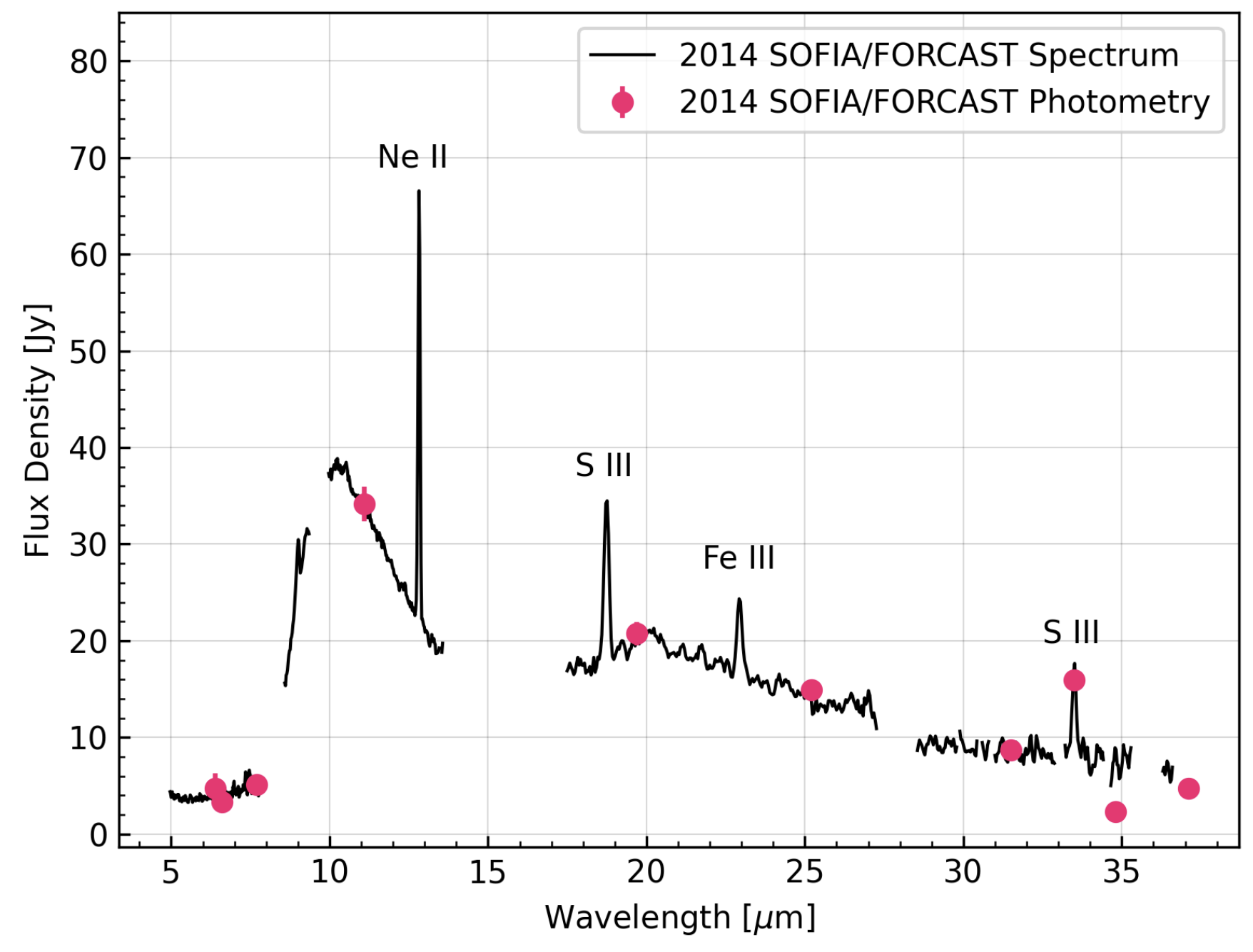}
    \caption{SOFIA multi-wavelength spectrum (solid black line) and photometry (filled pink circles). The spectrum is the data after telluric correction. Gaps in the spectrum across wavelength range are either due to wavelength regions not captured by the grisms or data filtered out during telluric correction. Line ionization \ states are labeled. The vertical pink lines overlaying the pink circles are the photometric error bars which, if not visible, are smaller than the plotting symbols. The continuum in both spectra is fueled by thermal emission from dust, and the significant feature at 10$\mu$m is likely due to silicate emission. There are no PAH emission lines at 7.7 or 11.3 $\mu$m.}
    \label{fig:sofia_spectrum}
\end{figure*}

RY Scuti is thought to be an immediate Wolf-Rayet binary progenitor system 
due to its significant mass loss and mass transfer \citep{antokhina88, giuricin81, smith02}.
In the near future, RY Scuti could evolve to resemble the very well studied $\gamma^{2}$ Vel 
system, with respective primary and secondary component masses of 9.0 $\pm$ 0.6~\Msun{} 
and 28.5 $\pm$ 1.1~\Msun{} \citep{north07}. In these systems, stripping of the donor star's outer envelope via RLOF will continue to reveal material in the star's inner envelope and core that is extremely hydrogen depleted. Later in the star's life, up to the days before it goes supernova, RLOF stripping will also reveal heavier metals such as sulfur and iron synthesized through oxygen and silicon burning \citep{grundstorm07}. The study of IR forbidden emission lines emitted by heavy metals in the ionized ring is 
therefore important. The 12.81 $\mu$m \ion{Ne}{2} line and other neon lines are especially important, as 
neon is the fourth most abundant element in WR stars, superseded only by oxygen, 
carbon, and helium \citep{smith05}. Determination of metal abundances indicative of stellar evolutionary stage via concurrent study of photoionization simulation and spectral 
analysis yields insight into the evolutionary stage of RY Scuti, and extends to later stage stellar remnants as WR systems themselves are progenitor systems of Type Ib/c supernovae \citep{tinyanont19}.

Here, we discuss a SOFIA (+FORCAST) study of RY Scuti. Section \ref{sec:obs} describes 
the observations, both spectroscopic and photometric, Section \ref{sec:sim} discusses the 
photoionization modelling (Cloudy; \citealt{chitzikos23}) of the RY Scuti IR spectrum and associated model parameters, 
Section \ref{sec:res_and_disc} summarizes the model results and their significance, and 
Section \ref{sec:future_con} reiterates this study's main conclusions.

\begin{table*}
\centering
\setlength{\tabcolsep}{14pt} 
\renewcommand{\arraystretch}{1.2} 
\caption{SOFIA Integrated Line Intensities\tablenotemark{a)} \label{tab:int_lines_sofia}}
\begin{tabular}{lcccc}
\hline
\hline
 & Wavelength & Flux & SNR & FWHM \\
Ion & ($\mu$m) & ($10^{-11}$ erg s$^{-1}$ cm$^{-2}$) &  & ($\mu$m) \\
\hline
\ion{Ne}{2} & 12.83 & 6.653 $\pm$ 0.224 & 79.0 & 0.076 \\
\ion{S}{3}  & 18.74 & 2.831 $\pm$ 0.152 & 29.7 & 0.18 \\
\ion{Fe}{3} & 22.94 & 1.001 $\pm$ 0.082 & 24.7 & 0.20 \\
\ion{S}{3}  & 33.50 & 0.381 $\pm$ 0.024 & 6.3 & 0.15 \\
\hline
\end{tabular}
\tablenotetext{a)\ }{The integrated IR line intensities derived from the SOFIA (+FORCAST) telluric-corrected spectrum (Figure~\ref{fig:sofia_spectrum}) using Gaussian fits. The wavelength, FWHM, and SNR of each line is also included.}
\end{table*}
\section{Observations} \label{sec:obs}
\subsection{Spectroscopy} \label{sec:spectrum}
A 5-37 $\mu$m spectrum of RY Scuti was obtained on 2014 June 06 
(program ID PID 02\_0101; \citealt{arneson18}) using the NASA Stratospheric Observatory for 
Infrared Astronomy's \citep[SOFIA,][]{young12} Faint Object infraRed 
CAmera instrument for the SOFIA Telescope \citep[FORCAST,][]{herter18}, a mid-infrared 
spectrograph and camera. Using the 4\farcs7 slit, all four FORCAST 
gratings were employed, G063 (4.95 to 7.85 $\mu$m), G111 (8.59 to 13.72 $\mu$m,), 
G227 (17.45 to 27.34 $\mu$m), and G329 (28.52 to 36.70 $\mu$m), with 
respective exposure times of 59.96s, 287.79s, 989.56s, and 1752.76s.
Level~3 processed SOFIA data were retrieved from the IRSA archive\footnote{\url{https://irsa.ipac.caltech.edu/}}, and are is presented in Figure~\ref{fig:sofia_spectrum}. Four forbidden transmission lines were 
detected, including three new lines (\ion{S}{3} at 18.71 $\mu$m, \ion{Fe}{3} at 22.92 $\mu$m, and \ion{S}{3} at 33.47 $\mu$m) not identified in previous RY Scuti spectra \citep{gehrz95}.

\begin{figure*}[!hbt]
    \centering
    \includegraphics[width=4in]{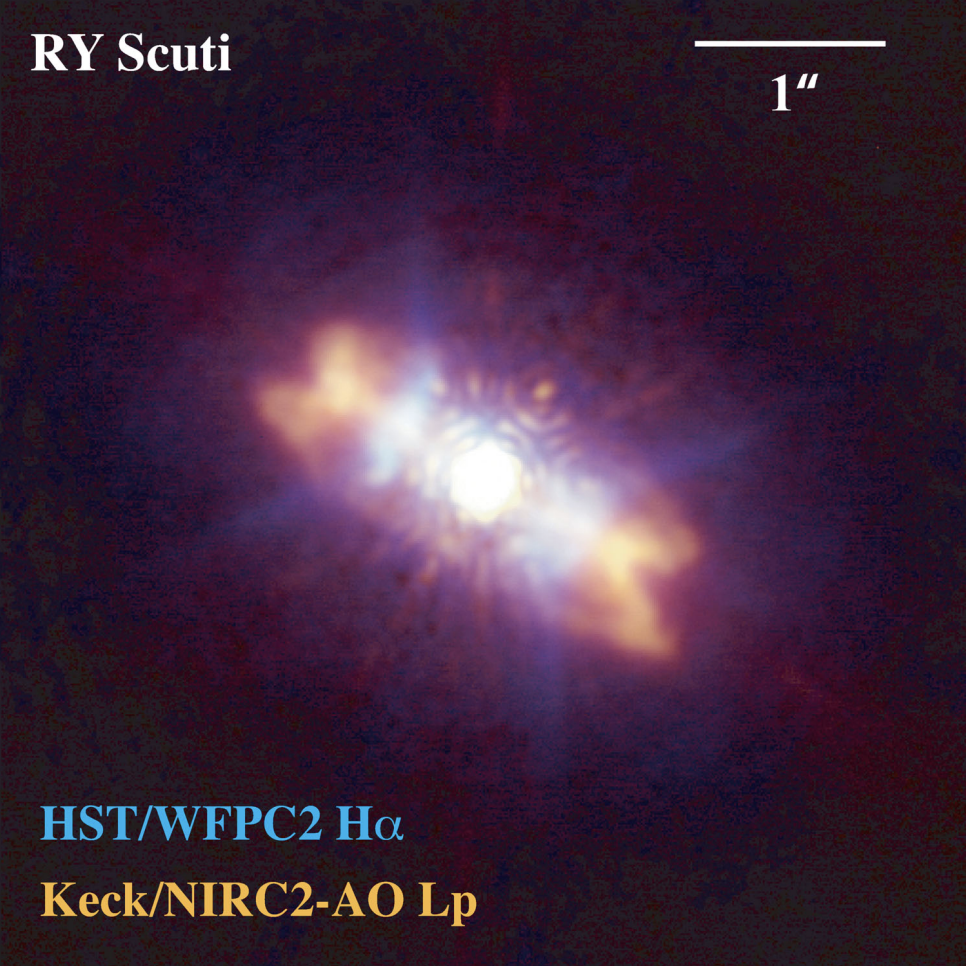}
    \caption{Reproduction of the color composite image of RY Scuti from \citet{smith11}. The blue portion is H$\alpha$ emission, the region used to determine the best fit height and inner and outer radii values for the Cloudy nebular model. The orange portion is IR emission from dust. This image illustrates that dust and gas emission originate in separate regions of the nebula.}
    \label{fig:halpha}
\end{figure*}

In accordance with SOFIA grism line analysis techniques, a telluric correction beyond standard pipeline processing was applied to the data to mask any points with transmission $\leq$ 0.75. The purpose of this additional masking was to exclude regions of the spectrum from model analysis that suffer from deep telluric absorption and thus a higher degree of uncertainty. This additional processing masked three other forbidden emission lines which are not presented in this work. While this reduced the total number of lines available for simulation and analysis, the remaining four lines have comparatively smaller uncertainties, leading to more reliable simulation results. The four remaining lines also represent all three elements observed in the original spectrum, so no element is excluded from this study due to telluric masking.

\subsection{Integrated Line Intensities} \label{sec:intline}

Integrated line intensities found in the SOFIA spectrum were calculated by employing Python-based Astropy Gaussian fitters
\citep{astropy13, astropy18, astropy22} 
to model the emission lines in the telluric corrected spectrum (Figure~\ref{fig:sofia_spectrum}).
Four individual Chebyshev polynomial continuum fit models were applied to and 
subtracted from each of the four wavelength regions observed by each of the four grisms. Integrated line intensities were calculated by fitting a Gaussian function to each continuum subtracted emission line and integrating under the area of each respective curve. Errors were determined using the associated covariance matricies, with the square root of each diagonalized matrix corresponding to the uncertainty in each line. The signal-to-noise ratio (SNR) and full-width-half-maximum (FWHM) of each line were calculated using Specutils \citep{specutils18}. Integrated flux values, corresponding uncertainties, line SNR, and line FWHM are all reported in Table \ref{tab:int_lines_sofia}.

\subsection{Photometry} \label{sec:photometry}
Contemporaneous FORCAST imaging observations of RY Scuti were obtained as part of PID 02\_0101. Images were obtained in ten narrow and board-band filters with effective wavelengths of 6.4, 6.6, 7.7, 11.1, 19.7, 25.2, 31.5, 33.5, 34.8, and 37.1 $\mu$m. Photometry of
RY Scuti was performed in effective circular diameters chosen to be $\simeq$ 3 times that of the cited FORCAST Cycle 7 point-spread function (PSF), with background aperture sizes 
$\simeq$ 2 times larger. Errors were determined by adding the relative aperture flux error, 
relative flux calibration error, uncertainty in the flux calibration model 
at the observed wavelength, and the aperture error in the variance plane in quadrature, and multiplying by the background subtracted 
aperture flux. Eight individual images were taken at wavelength 6.4 $\mu$m, so the 
final photometric value was calculated by averaging, with errors added in quadrature. The final photometric values and uncertainties are depicted as filled pink circles in in Figure \ref{fig:sofia_spectrum}, with error bars represented by vertical lines. Individual photometric wavelengths, values, and uncertainties are
reported in Table \ref{tab:photometry}.

\section{Photoionization Model} \label{sec:sim}
Given the presence of metallic forbidden lines in Figure~\ref{fig:sofia_spectrum}, 
the spectral synthesis photoionization code Cloudy \citep{ferland17, chitzikos23} was used for modeling.
Cloudy works by solving statistical, thermal, and chemical equilibrium equations based on a variety of input parameters which quantify both the gas cloud responsible for line emission and the radiation source responsible for gas cloud excitation. Cloudy has been used to simulate photoionization in many different types of objects both similar and dissimilar to RY Scuti, from the mid-IR emission of planetary nebulae \citep{gomez24}, to metallicities of systems found in the circumgalactic medium \citep{cristiani24}, to modeling the ejecta in classical novae across multiple epochs of their post-outburst 
evolution \citep{pandey22, habtie24}. 

In this study, the radiation source is the dual blackbody emission by both stars in the RY Scuti binary, and the line-emitting gas cloud is the inner, ionized region of RY Scuti's circumstellar nebula. Many model input parameters can be specified to characterize and quantify both the radiation source and emitting cloud, and those utilized in this study include effective blackbody temperature, blackbody luminosity, radially dependent cloud electron number density, cloud dust, cloud inner and outer radius, cloud height, cloud geometry, source distance, filling factor, covering factor, and cloud chemical abundances and depletion factors. For specific, quantified parameter values, see Section \ref{subsec:quant_params} and Table \ref{tab:observed}.

Initial models were created by exploring blackbody temperature, cloud electron number density, and filling factor values based off of canonical values and ranges found in the literature \citep{lang92, gehrz95, smith99, smith02, menshikov05, grundstorm07, djura08}. Exploration proceeded until the simulated integrated line intensities agreed with the observed integrated line intensities within an order of magnitude. Blackbody luminosity, cloud height, radius, geometry, covering factor, and source distance showed little variation in the literature and thus were chosen to remain static (i.e., without any initial experimentation). Models were assessed visually and those in clear disagreement with the observed SOFIA integrated line intensities were discarded and one or several parameter values were changed until agreement within an order of magnitude was reached.

\begin{figure*}[!hbt]
    \centering
    \includegraphics[width=7in]{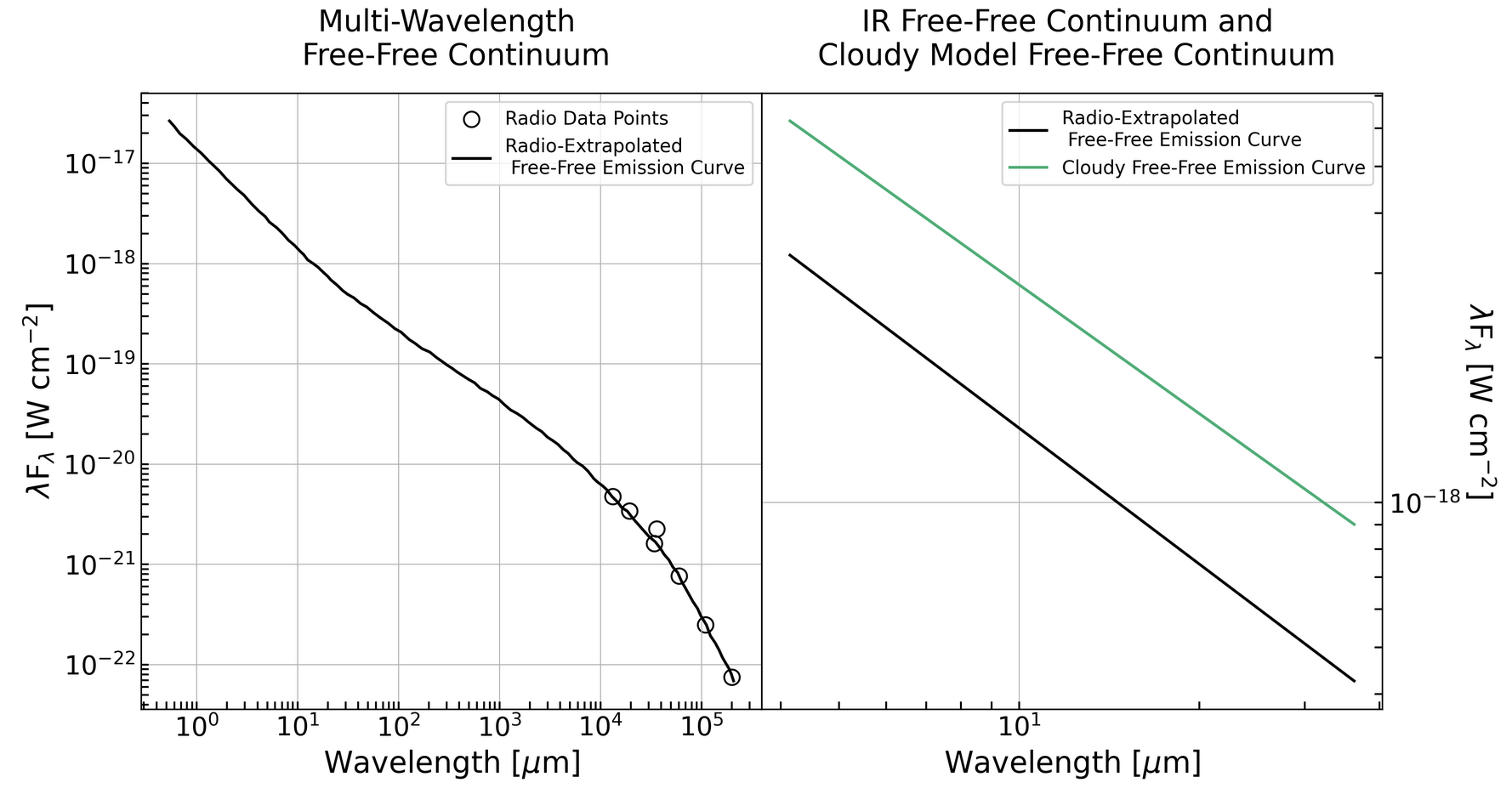}
    \caption{Left panel: Re-production of the radio data points and associated extrapolated thermal bremsstrahlung emission curve by a T=8000 K hydrogen nebula that becomes optically thick at 11cm from \citet{gehrz95}. The deviation of the optically thin free-free slope from a $\lambda^{-1}$ dependence at longer wavelengths is because the gaunt factor $g$ becomes large ($\simeq$ 5.5) at radio wavelengths. Right panel: The black line is the same free-free emission curve from the left panel shown only in the wavelength range explored in the Cloudy model. The green curve is the best fit power-law regression model of the free-free Cloudy continuum.}
    \label{fig:continuua}
\end{figure*}

Once order of magnitude agreement was reached between the synthetic spectrum and the observed spectrum, adjustments were made in the model helium abundance and hydrogen depletion to achieve order of magnitude agreement between the Cloudy model free-free continuum (green line in the right panel of Figure \ref{fig:continuua}) and the \citet{gehrz95} radio-extrapolated free-free continuum (black line in the left and right panels of Figure \ref{fig:continuua}). The model free-free continuum was resistant to changes in all other system parameters; adjusting the helium abundance and hydrogen depletion proved to be the only successful method in aligning the model free-free continuum and the \citet{gehrz95} radio-extrapolated continuum.

Once order of magnitude agreement was reached between the synthetic spectrum, the observed spectrum, and the associated free-free continua, adjustments were made in the chemical abundances of the three elements (neon, sulfur, and iron) responsible for producing the four observed forbidden transition lines. Changes in neon, sulfur, and iron abundance had no discernible affect on the appearance of the model free-free continuum. Chemical abundances were initially set to solar levels as determined by \citet{lodders21} and yielded models in poor agreement with the observed lines, as expected. Post-main sequence (PMS) values were also explored for non-free chemical abundances (i.e. elements other than helium, neon, sulfur, and iron) but showed no appreciable difference when compared with fiducial solar values and were therefore not used. Goodness of fit for each model spectrum was quantified by calculating a $\chi^{2}$ value via the following relation:

\begin{equation}
    \label{eq:1}
    \chi^{2} = \sum_{i=1}^{n}\frac{(M_{i} - O_{i})^{2}}{\sigma_{i}^{2}}
\end{equation}

\noindent where $M_{i}$ and $O_{i}$ respectively represent the 
modeled and observed integrated line intensities, and $\sigma_{i}$ represents the error in the observed integrated line intensities. The best model has the lowest $\chi^{2}$ value, and was obtained via a minimization routine employing the Powell method. In this routine, the best fit model is obtained for all lines simultaneously, with each line weighted equally. A reduced chi square was also considered for this analysis but was ultimately rejected due to the model's small number of degrees of freedom. Including each line's FWHM in the $\chi^{2}$ minimization was also considered, but was eventually rejected given the Cloudy's model's static (i.e. unchanging) resolution and therefore unchanging FWHM across varying chemical abundances. However, the observed and modeled FWHM values for each line are reported in Tables \ref{tab:int_lines_sofia} and \ref{tab:int_lines_cloudy} respectively.

One-$\sigma$ errors in the best fit chemical abundances were calculated by taking the standard deviation of the three ensembles of chemical abundance values iterated through during the $\chi^{2}$ minimization process. Errors in the best fit integrated line intensities were calculated using a Monte Carlo simulation approach. One hundred models were simulated for each line, with chemical abundance values drawn from a random normal distribution with the mean set to the best fit chemical abundance value of each species and the standard deviation set to the associated $1\sigma$ error of each species. Chemical abundances were the only free parameters in the MC simulation. The standard deviation of each of these simulated ensembles is the error in the corresponding best fit integrated line intensity. The best fit chemical abundances, associated errors, and $\chi^{2}$ value are reported in Table \ref{tab:observed}. The best fit model integrated line intensities, associated errors, adjacent free-free continuum, and line FWHM values are reported in Table \ref{tab:int_lines_cloudy} and can be seen in Figure \ref{fig:final_lines}.

\begin{deluxetable}{@{\extracolsep{14pt}}lcc}
\tablenum{2}
\setlength{\tabcolsep}{6pt} 
\tablecaption{Photometry Summary\tablenotemark{b)}} 
\label{tab:photometry}
\tablehead{
\colhead{Wavelength} & \colhead{Flux} \\
\colhead{($\mu$m)} & \colhead{(Jy)} 
}
\startdata
6.4 & 4.74 $\pm$ 0.79 \\
6.6 & 3.35 $\pm$ 0.11 \\
7.7 & 5.12 $\pm$ 0.16 \\
11.1 & 34.18 $\pm$ 0.89 \\
19.7 & 20.74 $\pm$ 0.61 \\  
25.2 & 14.91 $\pm$ 0.47 \\
31.5 & 8.72 $\pm$ 0.28 \\
33.5 & 15.93 $\pm$ 0.47 \\
34.8 & 2.34 $\pm$ 0.21 \\
37.1 & 4.76 $\pm$ 0.23 \\
\enddata
\tablenotetext{b)\ }{\, Summary of photometry values obtained alongside the newly reported RY Scuti spectrum in Figure \ref{fig:sofia_spectrum}.}
\end{deluxetable}

\begin{deluxetable}{@{\extracolsep{14pt}}lc}
\tablenum{3}
\setlength{\tabcolsep}{6pt} 
\tablecaption{Best fitting Cloudy model parameters and accompanying $\chi^{2}$ value. \label{tab:observed}}
\tablehead{
\colhead{Parameters} & \colhead{Entry}
}
\startdata
Blackbody temperature [K] & 27,000 \\
Luminosity [$\times$ 10$^{40}$ ergs] & 1.36 \\
Initial electron number density [$\times$ 10$^{6}$ cm$^{-3}$] & 1 \\
Inner radius [au] & 460 \\
Outer radius [au] & 1385 \\
Semi height [au] & 1360 \\
Distance [kpc] & 1.8 \\
Geometry & Cylindrical \\
Covering factor & 0.441 \\
Filling factor & 0.5 \\
Manually depleted hydrogen & 0.708 \\
Dust grains & None \\
Winds & None \\
Helium Abundance & -0.96 $\pm$ 0.0421 \\
Neon Abundance & -3.19 $\pm$ 0.0251 \\
Sulfur Abundance & -3.76 $\pm$ 0.0487 \\
Iron Abundance & -2.23 $\pm$ 0.0286 \\
$\chi^{2}$ & 14.38 \\
\enddata
\end{deluxetable}

\subsection{Model Parameter Values} \label{subsec:quant_params}

A temperature range between 25,000 K to 40,000~K was initially explored 
in increments of 1,000~K in Cloudy to \ determine first order-of-magnitude 
agreement between the synthetic and observed integrated line intensities.
The most common temperature value cited for the primary star is 38,000~K 
\citep{djura08, grundstorm07, lang92, smith99}, and the most common temperature 
value cited for the secondary star is 30,000~K \citep{djura08, grundstorm07, smith99}. The aggregate temperature value employed in this work for the synthetically 
combined (i.e., singular simulated blackbody irradiating system) is 27,000~K, on the lower end of most values typically employed in the literature. However, this value is not unprecedented as \citet{menshikov05} use
an identical value in their 2D-radiative transfer model of RY Scuti.

The bolometric luminosity of 1.36 $\times$ 10$^{40}$~ergs is one of the static model parameters (i.e., not changed or treated as a free parameter), and is adopted following that cited in \citet{gehrz95}. 

Since the radius of the inner, ionized ring likely corresponds to the system's Str\"omgren radius, all of the modeled hydrogen gas is assumed to be ionized. Cloudy is capable of modeling the electron number density as either radially constant or radially dependent. In the case of radial dependency, the number density is defined as:

\begin{equation}
    \label{eq:2}
    \frac{n(r)}{n(r_{0})} = \left(\frac{r}{r_{0}}\right)^{\alpha}  \mathrm{cm}^{-3}
\end{equation}

\noindent where $\alpha$ denotes the exponent of the power law, n(r) is the number density at a given radius, r, and n(r$_{0}$) is the number density at the inner radius of the cloud, r$_{0}$. Given that the toroidal nebula was created by a previous mass ejection event \citep{grundstorm07, smith11}, and that the resulting mass in the nebula is unlikely to be uniformly distributed as a result of this expulsion, radial dependence is assumed for the system's electron density. The corresponding power law exponent is chosen to be $\alpha$ = -3 to ensure a steady mass per unit volume in the model shell (i.e., $\dot{M}$ = constant).

The best fit value for the electron density at the ring's inner radius, r$_{0}$, is 1 $\times$ 10$^{6}$ cm$^{-3}$. While this is 5 times larger than the electron density of 2 $\times$ 10$^{5}$ cm$^{-3}$ cited in \citet{smith02}, the steep power law dependence allows the electron density to reach the \citet{smith02} value at a radius of $\simeq$ 790 au, or $\simeq$ $\frac{1}{3}$ of the way to the outer edge of the nebula.

Cylindrical geometry is assumed for the nebula, and inner radius, outer radius, and height, which define the boundaries of the cylindrical model, are adopted via visual inspection and trigonometric calculation from H$\alpha$ emission seen in the RY Scuti color composite image in Figure \ref{fig:halpha}. The inner shell radius is found to be 460 au and the outer radius is found to be 1385 au, similar to other values in the literature \citep{gehrz95, smith01, grundstorm07, smith11}. The cloud's linear projected extent (i.e., height) is found to be 1360 au.

The expansion parallax (see \citealt{mellema04}) of the ionized gas shell gives a distance of 1.8 $\pm$ 0.1 kpc, and the GAIA parallax gives a distance of 2.1 $\pm$ 0.1 kpc.  A distance of 1.8 kpc is assumed since our calculations refer to observations of the expanding gas shell, in agreement with past literature values \citep{king79, smith01, smith02}.

Given that RY Scuti is a PMS star exhibiting RLOF, and a likely WR progenitor, the system is hydrogen depleted. As such, the modeled hydrogen was manually depleted, and the best fit value is 0.708, or depletion by 29.2\%.

A covering factor was found by determining the cylindrical model's fractional projected surface area coverage on the sky. For simplicity, this area was calculated via:

\begin{equation}
    \label{eq:3}
    \Omega = \frac{4 \pi H}{\sqrt{H^{2} + 4R^{2}}}     \:\: \mathrm{sr}
\end{equation}

\noindent where H is the cylinder's height and R is the cylinder's outer radius. This value was then divided by 4$\pi$ steradians to determine the fractional coverage of 0.441, which is adopted as the covering factor value.

A canonical filling factor value of 0.5 is adopted to represent the not entirely smooth nor entirely clumpy distribution of gas in the inner, ionized nebula, as seen in Figure \ref{fig:halpha}. While it's possible this value accurately describes the degree of clumpiness in the system's gas, degeneracy between filling factor and covering factor values could also mean the filling factor is accidentally capturing the nebula's ring structure. As illustrated by \citet{smith99}, the ring structure in the gas precludes the nebula from covering the full extent of the cylinder's projected surface area as suggested by the simplified model covering factor value of 0.441. This means the true filling factor could be higher than 0.5 (i.e., smoother gas), and the true covering factor could be lower than 0.441 (i.e., smaller fractional coverage).

Given that only the inner, ionized region of the nebula is modeled, and dust and gas emission originate in separate regions of the nebula as seen in Figure \ref{fig:halpha}, no dust grains are included in the Cloudy model.

FORCAST gratings G063, G227, and G329 have a spectral resolution value of $R = 128$, with grating G111 at $R = 151$, resulting in resolvable velocities of $v = 2343 $ km s$^{-1}$ for gratings G063, G227, and G329, and $v = 1986 $ km s$^{-1}$ for grating G111. Expansion velocities of the inner, ionized ring are reported from 42 km s$^{-1}$ up to $\simeq$ 1000 km s$^{-1}$ at the nebula's poles \citep{smith11}, both notably smaller velocities than that which can be resolved in the spectrum. Stellar outflow (i.e., wind) is therefore not considered in the model.

\setcounter{table}{3}
\begin{table*}
\centering
\setlength{\tabcolsep}{14pt} 
\renewcommand{\arraystretch}{1.2} 
\caption{Cloudy Integrated Line Intensities\tablenotemark{c)} \label{tab:int_lines_cloudy}}
\begin{tabular}{lcccc}
\hline
\hline
 & Wavelength & Flux & Adjacent Continuum & FWHM \\
Ion & ($\mu$m) & ($10^{-11}$ erg s$^{-1}$ cm$^{-2}$) & ($10^{-11}$ erg s$^{-1}$ cm$^{-2}$) & ($\mu$m) \\
\hline
\ion{Ne}{2} & 12.8101 & 6.682 $\pm$ 0.618 & 2.260 & 0.045 \\
\ion{S}{3}  & 18.7078 & 3.141 $\pm$ 0.289 & 1.621 & 0.065 \\
\ion{Fe}{3} & 22.9190 & 0.957 $\pm$ 0.103 & 1.353 & 0.082 \\
\ion{S}{3}  & 33.4704 & 0.305 $\pm$ 0.028 & 0.971 & 0.12 \\
\hline
\end{tabular}
\tablenotetext{c)\ }{Best fit Cloudy model integrated line intensities with corresponding ion, wavelength, associated adjacent free-free continuum, and line FWHM values.}
\end{table*}

\section{Results and Discussion}
\label{sec:res_and_disc}
\begin{figure*}[!hbt]
    \centering
    \includegraphics[width=4in]{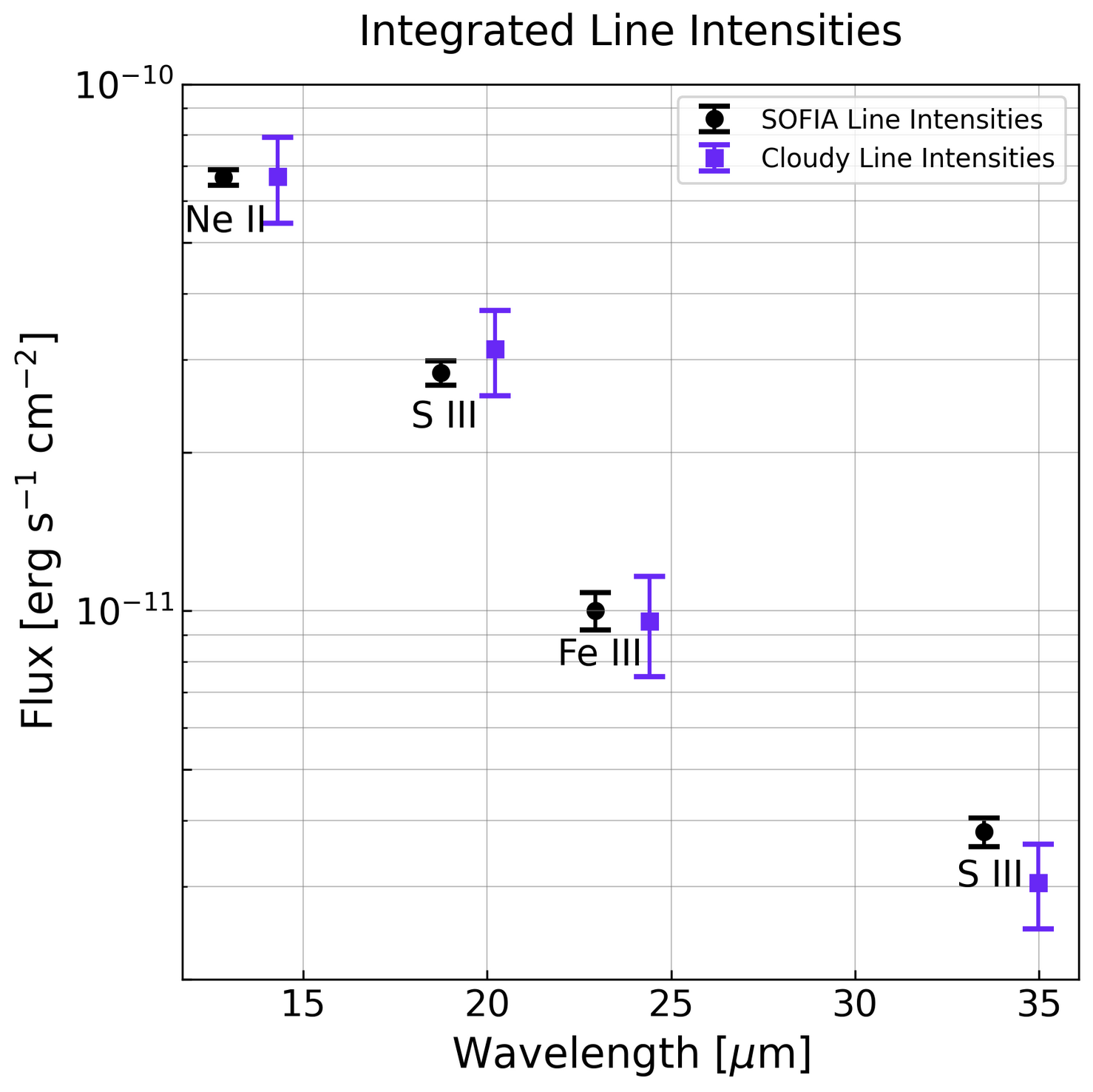}
    \caption{The black filled circles represent the SOFIA integrated line intensities and corresponding 1$\sigma$ uncertainties. The purple filled circles are the best fit Cloudy model integrated line intensities and uncertainties calculated via Monte Carlo simulation. Wavelength offsets between the SOFIA and Cloudy lines are artificial and have been introduced solely for visual comparison and clarity, with the true wavelength for each ion corresponding to the black filled circle (i.e. the SOFIA line intensities). All model integrated line intensities are within SOFIA errors.}
    \label{fig:final_lines}
\end{figure*}

Figure~\ref{fig:final_lines} summarizes the results of the best fitting Cloudy model alongside the SOFIA observations. The black filled circles and associated error bars are the four observed SOFIA integrated line intensity values, summarized in Table \ref{tab:int_lines_sofia}, and the purple filled squares and associated errors are the $\chi^{2}$ minimized (i.e., best fit) Cloudy model integrated line intensity values, summarized in Table \ref{tab:int_lines_cloudy}.

The simulated Cloudy line intensity values for all ions agree with the observed SOFIA line intensities of the same species within errors. The simulated \ion{S}{3} line at 33.4704 $\mu$m is most offset from its observed counterpart, and nearly in tension outside errors, with a percentage difference of 22\%. 
In spite of this, the sulfur chemical abundance value that results in both sulfur line intensities remains the best fit abundance (i.e., $\chi^{2}$ minimized).

Both sulfur lines at 18.7078 and 33.4704~$\mu$m are emission lines of the same ionization \ state, hence their intensities are both dependent on the same chemical abundance value. This linkage makes it impossible to increase one line flux value without increasing the other, or decrease one line flux value without decreasing the other. This posed a challenge given that the first simulated \ion{S}{3} emission line was consistently larger than the observed emission line, and the second simulated \ion{S}{3} emission line was consistently smaller than the observed emission line. Numerous attempts were made to improve the model by increasing the sulfur abundance so the resulting intensities balanced the positive offset of the first line with the negative offset of the second more evenly than the reported best fit abundance. While these simulations provided the desired result, the accompanying $\chi^{2}$ value was larger than the $\chi^{2}$ value for the best fit chemical abundances. Those models were therefore rejected.

The primary result from this model is that all of the best fitting log$\frac{N[x]}{N[H]}$ chemical abundance values are elevated above solar values, with neon at -3.19 $\pm$ 0.0251 compared to a solar value of -3.85, sulfur at -3.76 $\pm$ 0.0487 compared to a solar value of -4.85, iron at -2.23 $\pm$ 0.0286 compared to a solar value of -4.52, and helium at -0.96 $\pm$ 0.0421 compared to a solar value of -1.076 \citep{lodders21}. A summary of these values can be seen in Table \ref{tab:observed}.

Comparison of the Cloudy neon, sulfur, and iron abundances to fiducial \citet{lodders21} solar abundances reveals that the three elements are respectively 4.57, 12.30, and 194.98 times more abundant in the Cloudy model than in the solar model. Looking specifically at neon, \citet{gehrz95} propose an overabundance range with respect to solar fiducial values of 1.6-10. The 4.57 value falls within this range. Closer inspection reveals that \citet{gehrz95} reference a fiducial neon solar abundance of $\simeq$ -4.0 as cited by \citet{cameron82} and \citet{anders89}. The solar abundances listed in \citet{lodders21} were used for this study, which cites a fiducial solar neon abundance of -3.85. By comparing the best fit neon value of -3.19 with the $\simeq$ -4.0 value, neon proves overabundant by a factor of 6.45, still inside the \citet{gehrz95} proposed range but even higher.

Though at first glance, it may seem plausible to interpret the elevated neon, sulfur, and iron abundances as indicative of increased metallicity in the RY Scuti system, this is likely not the case. Neon is formed through helium capture by carbon and oxygen nuclei during the carbon-burning phase \citep{woosley02}. Sulfur and iron are produced later. Sulfur is created via the hydrostatic burning of neon which forms an oxygen convective core and produces elements up to ${}^{32}$S \citep{lucertini22}. Iron is formed during the hydrostatic silicon-burning phase where silicon and other lighter elements undergo fusion in the core, ultimately forming elements up to ${}^{56}$Fe \citep{golovatiyi92, raphael95}. All three of these processes occur within the late stages of a star's evolution, with the carbon burning phase lasting a few hundred years, neon burning lasting about a year, and silicon burning lasting only about a day. Given that these abundances are observed in a nebula ejected approximately 100 years ago, and that RY Scuti has not yet gone supernova, it is unlikely that RY Scuti has created significant amounts of neon, let alone experienced oxygen or silicon burning.

Additionally, all abundances are listed relative to hydrogen (log($\frac{X}{H}$)), meaning the hydrogen depletion factor also plays a role in explaining the overabundance of these three metals. The best fit helium abundance found to align the free-free continua is overabundant relative to fiducial solar levels by a factor of 1.31. Since helium is produced via hydrogen burning, the system is almost certainly hydrogen-depleted, a conclusion corroborated by the Cloudy model's 0.708 best fit hydrogen depletion value. While this depletion indeed contributes to boosting the observed elevated chemical abundances, the observed elevated abundances, especially of iron, cannot by explained by hydrogen depletion alone since the metals are elevated by factors that are both different from one another and larger than the factor by which hydrogen is depleted. Some enrichment factor other than in situ nucleosynthesis must be at play. Thus, given its helium enhancement, hydrogen depletion, elevated metallic abundances, and continuous RLOF \citep{giuricin81, antokhina88, smith02, smith11}, RY Scuti remains a strong candidate for a WR progenitor system.

While these results affirm past postulations about RY Scuti's evolutionary stage, they are not infallible. The first avenue for improvement is exploring more robust mechanisms to constrain the hydrogen depletion and helium abundance. The bootstrapping from the radio free-free flux and comparing to IR emission lines is influenced by the relative helium abundance and hydrogen depletion factor, but may also be sensitive to other factors like ionization, temperature, or density inhomogeneities in the emission gas. Hence, over-interpretation of the derived helium abundance and hydrogen depletion is not warranted. One other possible avenue for improvement is the fact that the chemical abundances for neon and iron are determined based only on the integrated intensity of one spectral line, and sulfur based only on two. While the SOFIA spectrum used for this work precludes analysis of any other lines, making claims about chemical abundances in the system based only on one or two lines of the corresponding ionization state is not the most robust avenue for deriving conclusions. Future spectral analysis of RY Scuti should attempt to observe more forbidden transition lines of the elements in question. Other similar avenues for future work could include assessing emission lines from ionization states other than neon, sulfur, and iron, particularly those important to WR systems such as carbon and hydrogen; modeling RY Scuti using different photoionization software and cross checking with Cloudy; modeling the stars in RY Scuti as two individual blackbody sources as opposed to one singular source; and using a more robust fitting technique other than $\chi^{2}$ minimization, such as log likelihood. 



\section{Summary} \label{sec:future_con}

By simulating archival SOFIA (+FORCAST) spectral lines of late-stage binary RY Scuti with photoionization software Cloudy, chemical abundances in the nebula's inner, ionized ring were determined. Respective values for the helium, neon, sulfur, and iron abundances are -0.96 $\pm$ 0.0421, -3.19 $\pm$ 0.0251, -3.76 $\pm$ 0.0487, and -2.23 $\pm$ 0.0286, with helium, neon, sulfur, and iron respectively 1.31, 4.57, 12.30, and 194.98 times more abundant than fiducial solar levels. This, in tandem with hydrogen depletion and ongoing RLOF, suggests that RY Scuti is a valid theorized WR progenitor system, caught at just the right time in its evolution to serve as an important source of study for astrophysical knowledge of stellar evolution.

\begin{acknowledgments}
This research has made use of the NASA/IPAC Infrared Science Archive, which is funded by the National Aeronautics and Space Administration and operated by the California Institute of Technology. Based in part on observations made with the NASA/DLR Stratospheric Observatory for Infrared Astronomy (SOFIA). SOFIA is jointly operated by the Universities Space Research Association, Inc. (USRA), under NASA contract NNA17BF53C , and the Deutsches SOFIA Institut (DSI) under DLR contract 50 OK 2002 to the University of Stuttgart. RDG was supported, in part, by the United States Air Force. SHT would also like to thank Lindsey Gordon, John Hamilton Miller Jr., and Derek Perera for useful discussion and suggestions regarding this work. 
\end{acknowledgments}

\facilities{SOFIA \citep{young12}, FORCAST \citep{herter18}} 
\software{Astropy \citep{astropy13, astropy18, astropy22},
Cloudy \citep{chitzikos23}} 

\bibliographystyle{aasjournal}
\bibliography{bibliography.bib}

\begin{thebibliography}{}
\expandafter\ifx\csname natexlab\endcsname\relax\def\natexlab#1{#1}\fi
\providecommand{\url}[1]{\href{#1}{#1}}
\providecommand{\dodoi}[1]{doi:~\href{http://doi.org/#1}{\nolinkurl{#1}}}
\providecommand{\doeprint}[1]{\href{http://ascl.net/#1}{\nolinkurl{http://ascl.net/#1}}}
\providecommand{\doarXiv}[1]{\href{https://arxiv.org/abs/#1}{\nolinkurl{https://arxiv.org/abs/#1}}}

\bibitem[{{Anders} \& {Grevesse}(1989)}]{anders89}
{Anders}, E., \& {Grevesse}, N. 1989, \gca, 53, 197

\bibitem[{{Antokhina} \& {Cherepashchuk}(1988)}]{antokhina88}
{Antokhina}, E.~A., \& {Cherepashchuk}, A.~M. 1988, Soviet Astronomy Letters,
  14, 105

\bibitem[{{Arneson} {et~al.}(2018){Arneson}, {Shenoy}, {Smith}, \&
  {Gehrz}}]{arneson18}
{Arneson}, R.~A., {Shenoy}, D., {Smith}, N., \& {Gehrz}, R.~D. 2018, \apj, 864,
  31

\bibitem[{{Astropy Collaboration} {et~al.}(2013){Astropy Collaboration},
  {Robitaille}, {Tollerud}, {Greenfield}, {Droettboom}, {Bray}, {Aldcroft},
  {Davis}, {Ginsburg}, {Price-Whelan}, {Kerzendorf}, {Conley}, {Crighton},
  {Barbary}, {Muna}, {Ferguson}, {Grollier}, {Parikh}, {Nair}, {Unther},
  {Deil}, {Woillez}, {Conseil}, {Kramer}, {Turner}, {Singer}, {Fox}, {Weaver},
  {Zabalza}, {Edwards}, {Azalee Bostroem}, {Burke}, {Casey}, {Crawford},
  {Dencheva}, {Ely}, {Jenness}, {Labrie}, {Lim}, {Pierfederici}, {Pontzen},
  {Ptak}, {Refsdal}, {Servillat}, \& {Streicher}}]{astropy13}
{Astropy Collaboration}, {Robitaille}, T.~P., {Tollerud}, E.~J., {et~al.} 2013,
  \aap, 558, A33

\bibitem[{{Astropy Collaboration} {et~al.}(2018){Astropy Collaboration},
  {Price-Whelan}, {Sip{\H{o}}cz}, {G{\"u}nther}, {Lim}, {Crawford}, {Conseil},
  {Shupe}, {Craig}, {Dencheva}, {Ginsburg}, {VanderPlas}, {Bradley},
  {P{\'e}rez-Su{\'a}rez}, {de Val-Borro}, {Aldcroft}, {Cruz}, {Robitaille},
  {Tollerud}, {Ardelean}, {Babej}, {Bach}, {Bachetti}, {Bakanov}, {Bamford},
  {Barentsen}, {Barmby}, {Baumbach}, {Berry}, {Biscani}, {Boquien}, {Bostroem},
  {Bouma}, {Brammer}, {Bray}, {Breytenbach}, {Buddelmeijer}, {Burke},
  {Calderone}, {Cano Rodr{\'\i}guez}, {Cara}, {Cardoso}, {Cheedella}, {Copin},
  {Corrales}, {Crichton}, {D'Avella}, {Deil}, {Depagne}, {Dietrich}, {Donath},
  {Droettboom}, {Earl}, {Erben}, {Fabbro}, {Ferreira}, {Finethy}, {Fox},
  {Garrison}, {Gibbons}, {Goldstein}, {Gommers}, {Greco}, {Greenfield},
  {Groener}, {Grollier}, {Hagen}, {Hirst}, {Homeier}, {Horton}, {Hosseinzadeh},
  {Hu}, {Hunkeler}, {Ivezi{\'c}}, {Jain}, {Jenness}, {Kanarek}, {Kendrew},
  {Kern}, {Kerzendorf}, {Khvalko}, {King}, {Kirkby}, {Kulkarni}, {Kumar},
  {Lee}, {Lenz}, {Littlefair}, {Ma}, {Macleod}, {Mastropietro}, {McCully},
  {Montagnac}, {Morris}, {Mueller}, {Mumford}, {Muna}, {Murphy}, {Nelson},
  {Nguyen}, {Ninan}, {N{\"o}the}, {Ogaz}, {Oh}, {Parejko}, {Parley}, {Pascual},
  {Patil}, {Patil}, {Plunkett}, {Prochaska}, {Rastogi}, {Reddy Janga},
  {Sabater}, {Sakurikar}, {Seifert}, {Sherbert}, {Sherwood-Taylor}, {Shih},
  {Sick}, {Silbiger}, {Singanamalla}, {Singer}, {Sladen}, {Sooley},
  {Sornarajah}, {Streicher}, {Teuben}, {Thomas}, {Tremblay}, {Turner},
  {Terr{\'o}n}, {van Kerkwijk}, {de la Vega}, {Watkins}, {Weaver}, {Whitmore},
  {Woillez}, {Zabalza}, \& {Astropy Contributors}}]{astropy18}
{Astropy Collaboration}, {Price-Whelan}, A.~M., {Sip{\H{o}}cz}, B.~M., {et~al.}
  2018, \aj, 156, 123

\bibitem[{{Astropy Collaboration} {et~al.}(2022){Astropy Collaboration},
  {Price-Whelan}, {Lim}, {Earl}, {Starkman}, {Bradley}, {Shupe}, {Patil},
  {Corrales}, {Brasseur}, {N{\"o}the}, {Donath}, {Tollerud}, {Morris},
  {Ginsburg}, {Vaher}, {Weaver}, {Tocknell}, {Jamieson}, {van Kerkwijk},
  {Robitaille}, {Merry}, {Bachetti}, {G{\"u}nther}, {Aldcroft},
  {Alvarado-Montes}, {Archibald}, {B{\'o}di}, {Bapat}, {Barentsen},
  {Baz{\'a}n}, {Biswas}, {Boquien}, {Burke}, {Cara}, {Cara}, {Conroy},
  {Conseil}, {Craig}, {Cross}, {Cruz}, {D'Eugenio}, {Dencheva}, {Devillepoix},
  {Dietrich}, {Eigenbrot}, {Erben}, {Ferreira}, {Foreman-Mackey}, {Fox},
  {Freij}, {Garg}, {Geda}, {Glattly}, {Gondhalekar}, {Gordon}, {Grant},
  {Greenfield}, {Groener}, {Guest}, {Gurovich}, {Handberg}, {Hart},
  {Hatfield-Dodds}, {Homeier}, {Hosseinzadeh}, {Jenness}, {Jones}, {Joseph},
  {Kalmbach}, {Karamehmetoglu}, {Ka{\l}uszy{\'n}ski}, {Kelley}, {Kern},
  {Kerzendorf}, {Koch}, {Kulumani}, {Lee}, {Ly}, {Ma}, {MacBride}, {Maljaars},
  {Muna}, {Murphy}, {Norman}, {O'Steen}, {Oman}, {Pacifici}, {Pascual},
  {Pascual-Granado}, {Patil}, {Perren}, {Pickering}, {Rastogi}, {Roulston},
  {Ryan}, {Rykoff}, {Sabater}, {Sakurikar}, {Salgado}, {Sanghi}, {Saunders},
  {Savchenko}, {Schwardt}, {Seifert-Eckert}, {Shih}, {Jain}, {Shukla}, {Sick},
  {Simpson}, {Singanamalla}, {Singer}, {Singhal}, {Sinha}, {Sip{\H{o}}cz},
  {Spitler}, {Stansby}, {Streicher}, {{\v{S}}umak}, {Swinbank}, {Taranu},
  {Tewary}, {Tremblay}, {de Val-Borro}, {Van Kooten}, {Vasovi{\'c}}, {Verma},
  {de Miranda Cardoso}, {Williams}, {Wilson}, {Winkel}, {Wood-Vasey}, {Xue},
  {Yoachim}, {Zhang}, {Zonca}, \& {Astropy Project Contributors}}]{astropy22}
{Astropy Collaboration}, {Price-Whelan}, A.~M., {Lim}, P.~L., {et~al.} 2022,
  \apj, 935, 167

\bibitem[{{Cameron}(1982)}]{cameron82}
{Cameron}, A.~G.~W. 1982, in Essays in Nuclear Astrophysics, ed. C.~A.
  {Barnes}, D.~D. {Clayton}, \& D.~N. {Schramm}, 23

\bibitem[{{Chatzikos} {et~al.}(2023){Chatzikos}, {Bianchi}, {Camilloni},
  {Chakraborty}, {Gunasekera}, {Guzm{\'a}n}, {Milby}, {Sarkar}, {Shaw}, {van
  Hoof}, \& {Ferland}}]{chitzikos23}
{Chatzikos}, M., {Bianchi}, S., {Camilloni}, F., {et~al.} 2023, \rmxaa, 59, 327

\bibitem[{{Cristiani} {et~al.}(2024){Cristiani}, {Cupani}, {Trost},
  {D'Odorico}, {Guarneri}, {Lo Curto}, {Meneghetti}, {Di Marcantonio}, {Faria},
  {Gonz{\'a}lez Hern{\'a}ndez}, {Lovis}, {Martins}, {Milakovi{\'c}}, {Molaro},
  {Murphy}, {Nunes}, {Pepe}, {Rebolo}, {Santos}, {Schmidt}, {Sousa},
  {Sozzetti}, \& {Zapatero Osorio}}]{cristiani24}
{Cristiani}, S., {Cupani}, G., {Trost}, A., {et~al.} 2024, \mnras, 528, 6845

\bibitem[{{Djura{\v{s}}evi{\'c}} {et~al.}(2008){Djura{\v{s}}evi{\'c}}, {Vince},
  \& {Atanackovi{\'c}}}]{djura08}
{Djura{\v{s}}evi{\'c}}, G., {Vince}, I., \& {Atanackovi{\'c}}, O. 2008, \aj,
  136, 767

\bibitem[{{Djura{\v{s}}evi{\'c}} {et~al.}(2001){Djura{\v{s}}evi{\'c}},
  {Zakirov}, {Eshankulova}, \& {Erkapi{\'c}}}]{djura01}
{Djura{\v{s}}evi{\'c}}, G., {Zakirov}, M., {Eshankulova}, M., \& {Erkapi{\'c}},
  S. 2001, \aap, 374, 638

\bibitem[{{Edlen} \& {Swings}(1939)}]{edlen39}
{Edlen}, B., \& {Swings}, P. 1939, The Observatory, 62, 234

\bibitem[{{Ferland} {et~al.}(2017){Ferland}, {Chatzikos}, {Guzm{\'a}n},
  {Lykins}, {van Hoof}, {Williams}, {Abel}, {Badnell}, {Keenan}, {Porter}, \&
  {Stancil}}]{ferland17}
{Ferland}, G.~J., {Chatzikos}, M., {Guzm{\'a}n}, F., {et~al.} 2017, \rmxaa, 53,
  385

\bibitem[{{Gehrz} {et~al.}(2001){Gehrz}, {Smith}, {Jones}, {Puetter}, \&
  {Yahil}}]{gehrz01}
{Gehrz}, R.~D., {Smith}, N., {Jones}, B., {Puetter}, R., \& {Yahil}, A. 2001,
  \apj, 559, 395

\bibitem[{{Gehrz} {et~al.}(1995){Gehrz}, {Hayward}, {Houck}, {Miles},
  {Hjellming}, {Jones}, {Woodward}, {Prentice}, {Forrest}, {Libonate}, \&
  {Solomon}}]{gehrz95}
{Gehrz}, R.~D., {Hayward}, T.~L., {Houck}, J.~R., {et~al.} 1995, \apj, 439, 417

\bibitem[{{Giuricin} \& {Mardirossian}(1981)}]{giuricin81}
{Giuricin}, G., \& {Mardirossian}, F. 1981, \aap, 101, 138

\bibitem[{{Golovatyi} \& {Skulskii}(1992)}]{golovatiyi92}
{Golovatyi}, V.~V., \& {Skulskii}, M.~Y. 1992, \sovast, 36, 550

\bibitem[{{G{\'o}mez-Mu{\~n}oz} {et~al.}(2024){G{\'o}mez-Mu{\~n}oz},
  {Garc{\'\i}a-Hern{\'a}ndez}, {Barzaga}, {Manchado}, \&
  {Huertas-Rold{\'a}n}}]{gomez24}
{G{\'o}mez-Mu{\~n}oz}, M.~A., {Garc{\'\i}a-Hern{\'a}ndez}, D.~A., {Barzaga},
  R., {Manchado}, A., \& {Huertas-Rold{\'a}n}, T. 2024, \aap, 682, L18

\bibitem[{{Grundstrom} {et~al.}(2007){Grundstrom}, {Gies}, {Hillwig},
  {McSwain}, {Smith}, {Gehrz}, {Stahl}, \& {Kaufer}}]{grundstorm07}
{Grundstrom}, E.~D., {Gies}, D.~R., {Hillwig}, T.~C., {et~al.} 2007, \apj, 667,
  505

\bibitem[{{Habtie} {et~al.}(2024){Habtie}, {Das}, {Pandey}, {Ashok}, \&
  {Dubovsky}}]{habtie24}
{Habtie}, G.~R., {Das}, R., {Pandey}, R., {Ashok}, N.~M., \& {Dubovsky}, P.~A.
  2024, \mnras, 527, 1405

\bibitem[{{Herter} {et~al.}(2018){Herter}, {Adams}, {Gull}, {Schoenwald},
  {Keller}, {Pirger}, {Henderson}, {Stacey}, {Nikola}, {De Buizer}, {Vacca}, \&
  {Ennico}}]{herter18}
{Herter}, T.~L., {Adams}, J.~D., {Gull}, G.~E., {et~al.} 2018, Journal of
  Astronomical Instrumentation, 7, 1840005

\bibitem[{{Hix}(1995)}]{raphael95}
{Hix}, W.~R. 1995, PhD thesis, Harvard University, Massachusetts

\bibitem[{{King} \& {Jameson}(1979)}]{king79}
{King}, A.~R., \& {Jameson}, R.~F. 1979, \aap, 71, 326

\bibitem[{{Lang}(1992)}]{lang92}
{Lang}, K.~R. 1992, {Astrophysical Data I. Planets and Stars.}

\bibitem[{{Lodders}(2021)}]{lodders21}
{Lodders}, K. 2021, \ssr, 217, 44

\bibitem[{{Lucertini} {et~al.}(2022){Lucertini}, {Monaco}, {Caffau},
  {Bonifacio}, \& {Mucciarelli}}]{lucertini22}
{Lucertini}, F., {Monaco}, L., {Caffau}, E., {Bonifacio}, P., \& {Mucciarelli},
  A. 2022, \aap, 657, A29

\bibitem[{{Melikian} {et~al.}(2010){Melikian}, {Tamazian}, {Docobo},
  {Karapetian}, {Kostandian}, \& {Samsonian}}]{melikian10}
{Melikian}, N.~D., {Tamazian}, V.~S., {Docobo}, J.~A., {et~al.} 2010,
  Astrophysics, 53, 202

\bibitem[{{Mellema}(2004)}]{mellema04}
{Mellema}, G. 2004, \aap, 416, 623

\bibitem[{{Men'shchikov} \& {Miroshnichenko}(2005)}]{menshikov05}
{Men'shchikov}, A.~B., \& {Miroshnichenko}, A.~S. 2005, \aap, 443, 211

\bibitem[{{North} {et~al.}(2007){North}, {Tuthill}, {Tango}, \&
  {Davis}}]{north07}
{North}, J.~R., {Tuthill}, P.~G., {Tango}, W.~J., \& {Davis}, J. 2007, \mnras,
  377, 415

\bibitem[{{Pandey} {et~al.}(2022){Pandey}, {Das}, {Shaw}, \&
  {Mondal}}]{pandey22}
{Pandey}, R., {Das}, R., {Shaw}, G., \& {Mondal}, S. 2022, \apj, 925, 187

\bibitem[{{Popper}(1943)}]{popper43}
{Popper}, D.~M. 1943, \apj, 97, 394

\bibitem[{Price-Whelan \& et~al.(2018)}]{specutils18}
Price-Whelan, A.~M., \& et~al. 2018, Specutils: A Python Package for
  Spectroscopic Analysis,  Zenodo

\bibitem[{{Sahade} {et~al.}(2002){Sahade}, {West}, \& {Skul'Sky}}]{sahade02}
{Sahade}, J., {West}, R.~M., \& {Skul'Sky}, M.~Y. 2002, \rmxaa, 38, 259

\bibitem[{{Skulskii}(1992)}]{skulskii92}
{Skulskii}, M.~Y. 1992, \sovast, 36, 411

\bibitem[{{Smith} \& {Houck}(2005)}]{smith05}
{Smith}, J.-D.~T., \& {Houck}, J.~R. 2005, \apj, 622, 1044

\bibitem[{{Smith} {et~al.}(2011){Smith}, {Gehrz}, {Campbell}, {Kassis}, {Le
  Mignant}, {Kuluhiwa}, \& {Filippenko}}]{smith11}
{Smith}, N., {Gehrz}, R.~D., {Campbell}, R., {et~al.} 2011, \mnras, 418, 1959

\bibitem[{{Smith} {et~al.}(2001){Smith}, {Gehrz}, \& {Goss}}]{smith01}
{Smith}, N., {Gehrz}, R.~D., \& {Goss}, W.~M. 2001, \aj, 122, 2700

\bibitem[{{Smith} {et~al.}(1999){Smith}, {Gehrz}, {Humphreys}, {Davidson},
  {Jones}, \& {Krautter}}]{smith99}
{Smith}, N., {Gehrz}, R.~D., {Humphreys}, R.~M., {et~al.} 1999, \aj, 118, 960

\bibitem[{{Smith} {et~al.}(2002){Smith}, {Gehrz}, {Stahl}, {Balick}, \&
  {Kaufer}}]{smith02}
{Smith}, N., {Gehrz}, R.~D., {Stahl}, O., {Balick}, B., \& {Kaufer}, A. 2002,
  \apj, 578, 464

\bibitem[{{Tinyanont} {et~al.}(2019){Tinyanont}, {Lau}, {Kasliwal}, {Maeda},
  {Smith}, {Fox}, {Gehrz}, {De}, {Jencson}, {Bally}, \& {Masci}}]{tinyanont19}
{Tinyanont}, S., {Lau}, R.~M., {Kasliwal}, M.~M., {et~al.} 2019, \apj, 887, 75

\bibitem[{{Walborn}(1982)}]{walborn82}
{Walborn}, N.~R. 1982, \aj, 87, 1300

\bibitem[{{Woosley} {et~al.}(2002){Woosley}, {Heger}, \& {Weaver}}]{woosley02}
{Woosley}, S.~E., {Heger}, A., \& {Weaver}, T.~A. 2002, Reviews of Modern
  Physics, 74, 1015

\bibitem[{{Young} {et~al.}(2012){Young}, {Becklin}, {Marcum}, {Roellig}, {De
  Buizer}, {Herter}, {G{\"u}sten}, {Dunham}, {Temi}, {Andersson}, {Backman},
  {Burgdorf}, {Caroff}, {Casey}, {Davidson}, {Erickson}, {Gehrz}, {Harper},
  {Harvey}, {Helton}, {Horner}, {Howard}, {Klein}, {Krabbe}, {McLean}, {Meyer},
  {Miles}, {Morris}, {Reach}, {Rho}, {Richter}, {Roeser}, {Sandell}, {Sankrit},
  {Savage}, {Smith}, {Shuping}, {Vacca}, {Vaillancourt}, {Wolf}, \&
  {Zinnecker}}]{young12}
{Young}, E.~T., {Becklin}, E.~E., {Marcum}, P.~M., {et~al.} 2012, \apjl, 749,
  L17

\end{thebibliography}


\end{document}